# Anisotropic electronic properties of *a*-axis-oriented $Sr_2IrO_4$ epitaxial thin-films

J. Nichols, O. B. Korneta, J. Terzic, L. E. De Long, G. Cao, J. W. Brill, and S. S. A. Seo[a]

*Department of Physics and Astronomy, University of Kentucky, Lexington, KY 40506, USA*

## Abstract

We have investigated the transport and optical properties along the *c*-axis of *a*-axis-oriented $Sr_2IrO_4$ epitaxial thin-films grown on $LaSrGaO_4$ (100) substrates. The *c*-axis resistivity is approximately one order of magnitude larger than that of the *ab*-plane. Optical absorption spectra with E⊥*c* polarization show both Ir $5d$ intersite transitions and charge-transfer transitions (O $2p$ to Ir $5d$), while E//*c* spectra show only the latter. The structural anisotropy created by biaxial strain in *a*-axis-oriented thin-films also changes the electronic structure and gap energy. These *a*-axis-oriented, epitaxial thin-films provide a powerful tool to investigate the highly anisotropic electronic properties of $Sr_2IrO_4$.

[a] E-mail: a.seo@uky.edu

A layered iridate compound, $Sr_2IrO_4$, which is an antiferromagnetic insulator ($T_N \sim 240$ K),[1,2] has recently attracted substantial interest due to its exotic electronic state. Coexisting electronic correlations and strong spin-orbit coupling of $5d$ electrons has led to the formation of the $J_{eff} = 1/2$ Mott state in $Sr_2IrO_4$ (SIO-214).[3] Even though debate continues concerning its ground state (i.e., Mott insulator vs. Slater insulator[4,5]), this compound has unprecedented potential for electronic device applications. For example, unconventional superconductivity is theoretically predicted in doped SIO-214[6] and the strong spin-orbit interaction is expected to result in novel electronic states such as topological insulators[7] and Weyl semimetals.[8,9] Recently, SIO-214 thin films have been grown and characterized,[10-12] resulting in a better understanding of the underlying physics of SIO-214 and providing impetus for developing device applications. However, only $c$-axis-oriented SIO-214 thin films have been synthesized thus far, which limits experimental access primarily to in-plane ($ab$-plane) properties. Hence, thin film studies of SIO-214 have produced results that are quite similar to those obtained for bulk SIO-214 crystals, whose naturally cleaved surfaces are also $ab$-planes. Since characterization of the $c$-axis of a number of layered oxides have revealed important physical information (e.g., the pseudo-gap energies in high-$T_c$ cuprates[13]), the fabrication of SIO-214 thin films with large $ac$-planes (or $bc$-planes) will permit investigations of important physical properties that are not readily accessible in typical bulk crystals and $c$-axis-oriented thin films.

In this Letter, we report the structural, transport, and optical properties of $a$-axis-oriented SIO-214 thin films, whose large surfaces ($5 \times 5$ mm$^2$) are $bc$-planes (or $ac$-planes). We have grown $a$-axis-oriented SIO-214 epitaxial thin films on $LaSrGaO_4$ (100) single-crystal substrates, where the [100]-direction is the surface-normal direction. In a similar study of layered $3d$ transition-metal oxides, $a$-axis-oriented thin films were grown on $LaSrAlO_4$ (100),[14] which has the same $K_2NiF_4$-structure as $LaSrGaO_4$ (LSGO). Due to differences in the tetragonal lattice parameters for SIO-214 (5.4979 Å and 25.798 Å)[15] and LSGO (3.852 Å and 12.68 Å),[16] the [110] and [$\bar{1}$10] directions of the SIO-214 thin films are parallel to the [100] and [010] directions of LSGO, respectively, and the thin film's $c$-axis lies parallel to the [001]-axis of the substrate, as schematically illustrated in Fig. 1. The [100], [010], and [001]-directions of the LSGO substrate are labeled as $a$, $b$, and $c$, respectively, and we use this notation in the following paragraphs.

Since the lattice mismatches between the substrate and the SIO-214 thin films are calculated as −0.92 % and −1.73 % along the $b$- and $c$-axes (Table 1), there is biaxial compressive strain along the $b$- and $c$- axes of the thin films. We have measured the transport and optical properties along the $ab$-plane and the $c$-axis of the SIO-214 thin films, which clearly show its anisotropic insulating nature. In particular, the $c$-axis optical spectrum has no absorption peaks except for the charge-transfer transition peak (from O $2p$ to Ir $5d$) above 2 eV. Our observation confirms that the low-energy optical transitions that exist near 0.5 eV and 1.0 eV in SIO-214 are due to *inter-site* optical transitions between Ir $5d$ orbitals that lie in the $ab$-plane.

We have grown $a$-axis-oriented, epitaxial SIO-214 thin films using a custom-built pulsed laser deposition system with *in-situ* reflection high-energy electron diffraction (RHEED) and *in-situ* optical spectroscopic ellipsometry.[17] Optimal growth parameters are oxygen partial pressure ($P_{O2}$) of 10 mTorr, substrate temperature of 700 ˚C, and laser (KrF excimer, $\lambda$ = 248 nm) fluence of 1.2 J/cm$^2$. We have monitored the thin film growth using RHEED, which shows a "layer-by-layer + island" growth mode, presumably due to the large surface energy of the film. A total film thickness of about 20 nm has been estimated by using 4 to 5 oscillations of the RHEED specular spot intensity during the initial growth.

The structure of these $a$-axis-oriented, epitaxial SIO-214 thin films has been identified using X-ray diffraction. The (220) and (440) thin-film peaks are only visible very near to the (200) and (400) substrate peaks in the $\theta$-$2\theta$ scan in Fig. 2 (a), ensuring that the films have an $a$-axis orientation. The FWHM of the rocking curves of the thin-film diffraction peaks are less than 0.07° (data not shown) suggesting that the samples have good crystallinity. In addition, the $bc$-plane epitaxy has been confirmed by pole figures and φ-scans (data not shown). In order to obtain lattice-strain information, X-ray reciprocal space maps have been measured near the (310) and (303)-reflections of the LSGO substrate for the $ab$- and $ac$-planes, respectively, as shown in Fig. 1(b) and 1(c). Note that there is biaxial compressive strain in the $bc$-plane resulting in the elongated $a$-axis of the SIO-214 thin film, even though strain relaxation easily occurs along the $b$-axis. The lattice parameters, lattice strain, and the Poison ratio are summarized in Table 1. It is noteworthy that artificial $ab$-plane anisotropy has been created by biaxial lattice stain in this sample geometry, i.e. the $a$-axis lattice parameter is longer than the $b$-axis. Therefore, the $a$-axis-oriented, epitaxial SIO-214 thin films have orthorhombic rather than tetragonal structure.

Figure 3 shows the electrical resistivity of the SIO-214 thin films for two current orientations: Samples were sliced and patterned into bar shapes to measure the temperature-dependent resistivity along the $b$-axis ($\rho_{ab}$) and $c$-axis ($\rho_c$) using conventional four-probe methods. Insulating behavior is clearly evident in both directions; however, the $c$-axis resistivity is about an order of magnitude larger than the $b$-axis resistivity. This anisotropy is also present in the Arrhenius plot shown in Fig. 3(b), where the dashed lines are fits to $\rho\,(T) = \rho_0 \exp(\Delta/2k_B T)$, where $\rho$, $\rho_0$, $\Delta$, and $k_B$ are the resistivity, proportionality constant, gap energy, and Boltzmann constant, respectively. Note that the values of $\Delta$ are estimated to be 97 meV (77 meV) at high temperature and 27 meV (24 meV) at low temperature for current applied along the $c$-axis ($b$- axis). A similar temperature-dependent behavior has been reported for bulk SIO-214 crystals, where the temperature dependence of the gap is primarily attributed to additional magnetic ordering below the magnetic transition temperature.[18]

Figure 4(a) shows optical absorption spectra of the SIO-214 thin films, which also exhibit anisotropy. The optical absorption coefficients are measured at room temperature with a Fourier-transform infrared spectrometer for energies in the range of $0.05 - 0.6$ eV, and a grating-type spectrometer for energies in the range of $0.5 - 6$ eV, using polarized incident light with E⊥$c$ or E//$c$. A schematic illustration of the measurement setup is presented in the inset of Fig 4(a), where LSGO and SIO-214 are blue and red, respectively. Two absorption peaks at around 0.5 eV ($\alpha$) and 1.0 eV ($\beta$) are clearly visible in the E⊥$c$ spectra, while no absorption peak is present at these energies in the E//$c$ spectra. The $\alpha$ and $\beta$ peaks have been already observed in the $ab$-plane of SIO-214 bulk crystals[19] and $c$-axis-oriented thin films[12]. They are interpreted as Ir $5d$ optical transitions between $J_{\text{eff}} = 1/2$ and $J_{\text{eff}} = 3/2$ states,[3,20] as schematically illustrated in Fig. 4(b). Note that the Ir $5d$ optical transitions reflect electron hopping between Ir sites in the $ab$-plane. The absence of the $\alpha$ and $\beta$ peaks in the E//$c$ optical spectrum confirms that inter-site optical transitions are forbidden in the E//$c$ polarization (Fig. 4(c)). However, both the E⊥$c$ and E//$c$ spectra exhibit a relatively isotropic feature at around 3 eV (A) due to charge-transfer optical transitions from O $2p$ states to Ir $5d$ states.

Note that the optical peak widths of the $\alpha$ and $\beta$ transitions in the E⊥$c$ spectrum are quite similar to those of SIO-214 thin films under isotropic $ab$-plane tensile strain with a compressed $c$-axis lattice. In the recent study of $c$-axis-oriented SIO-214 thin films (i.e., SIO-214 $c$-axis normal to substrate surface) deposited on various substrates with both *ab*-plane tensile and compressive strain,[12] the optical peak widths and positions exhibit a systematic dependence on lattice strain. The E⊥*c* optical spectrum (Fig. 4(a)) is similar to that for *c*-axis-oriented SIO-214 thin films grown on $SrTiO_3$ (100) and $GdScO_3$ (110) substrates, which are under isotropic tensile strain in the *ab*-plane with a decreased *c*-axis lattice parameter. Since there is *ab*-plane anisotropy in the *a*-axis oriented thin films discussed in this letter (Table 1), it is hard to draw a concrete conclusion concerning these data. However, the similarity of these two spectra suggests that changes in the *c*-axis lattice parameter (involving elongation or flattening of $IO_6$ octahedra) play an important role in changing the electronic structure of SIO-214. It is also noteworthy that an optical gap energy ($\Delta_{op}$) of about 0.2 eV is estimated from the onset of the E⊥*c* optical absorption spectrum, which is approximately two-thirds of the values obtained in Refs. [12] and [19]. This optical gap suppression, which is not observed in *c*-axis-oriented SIO-214 thin films, might be related to the *ab*-plane anisotropy in *a*-axis-oriented SIO-214 thin films. Since it is not clear how the artificial *ab*-plane anisotropy, i.e., strain-induced, different *a*- and *b*-axes lattice parameters, distorts the $IrO_6$ octahedra in *a*-axis-oriented thin films, microscopic characterizations such as scanning transmission electron microscopy[21,22] and resonant X-ray diffraction[23,24] will provide additional important information.

In summary, we have synthesized *a*-axis-oriented, epitaxial SIO-214 thin films with artificial *ab*-plane anisotropy and a flattened *c*-axis lattice on LSGO (100) substrates. We have observed that these thin films are insulating along both the *b*- and *c*-axes, but the *c*-axis resistivity is an order of magnitude larger than the *b*-axis resistivity. We have observed optical absorption spectra where the two optical peaks at 0.5 eV and 1.0 eV are only observed for E⊥*c*, which supports the view that these peaks originate from inter-site Ir $5d$ transitions. Since the large surface area ($bc$-plane) of these samples is orthogonal to the naturally cleaved surface of this compound, our sample geometry provides an important way to investigate the in-plane anisotropy and $c$-axis properties, which are not easily accessible in bulk crystals.

We appreciate Emily Bittle for her help with the FT-IR measurements. This research was supported by the NSF through Grant Nos. EPS-0814194 (the Center for Advanced Materials), DMR-1262261 (JWB), DMR-0856234 (GC), DMR-1265162 (GC), by U.S. DoE through Grant No. DE-FG02-97ER45653 (LED), and by the Kentucky Science and Engineering Foundation with the Kentucky Science and Technology Corporation through Grant Agreement No. KSEF-148-502-12-303 (SSAS).

Table 1. Lattice parameters and strain of $a$-axis-oriented epitaxial thin films of $Sr_2IrO_4$ (SIO-214).

| | Crystallographic direction | LSGO lattice parameters (Å) | SIO-214 film pseudo-cubic lattice parameters (Å) | Lattice mismatch (%)[*] | Lattice strain (%)[**] |
|---|---|---|---|---|---|
| $a$ | $[100]_{sub}$ // $[110]_{film}$ | 3.852 | 3.91 | —— | + 0.65 |
| $b$ | $[010]_{sub}$ // $[\bar{1}10]_{film}$ | 3.852 | 3.88 | − 0.92 | − 0.1 |
| $c$ | $[001]_{sub}$ // $[001]_{film}$ | 12.68 | 12.7 | − 1.73 | − 1.4 |

[*] Lattice mismatch is calculated from the pseudo-cubic lattice parameters of bulk SIO-214 ($d_{bulk}$) and substrates ($d_{sub}$) by $(d_{sub} - d_{bulk}) / d_{sub} \times 100$ (%).

[**] Lattice strain is estimated by using $\varepsilon = (d_{film} - d_{bulk}) / d_{bulk} \times 100$ (%).

Poisson's ratio, $v = \varepsilon_a / (\varepsilon_a - \varepsilon_b - \varepsilon_c) = 0.30$.

**Figure Captions**

FIG. 1  Schematic diagram of the sample geometry with the *a*-axis-oriented SIO-214 thin-film grown on the LSGO (100) substrate, where the $IrO_6$ octahedra are red and the Sr atoms are blue. The [110], [$\bar{1}$10], and [001] directions of the SIO-214 thin film are parallel to the *a*: [100], *b*: [010], and *c*: [001] directions of LSGO, respectively.  The large colored arrows represent the direction of crystal strain:  compressive (green) along the *b*- and *c*-axes and tensile (orange) along the *a*-axis.

FIG. 2  X-ray diffraction data for SIO-214 thin-films on LSGO:  (a) $\theta$-$2\theta$ scan of a thin film for which the (*ll*0) film peaks are clearly visible and confirm the *a*-axis orientation of the film. Reciprocal space maps of (b) the SIO-214 (420) film peak (black ×) near the LSGO (310) substrate peak (white cross), and (c) the SIO-214 (336) film peak (black ×) near the LSGO (303) substrate peak (white cross).

FIG. 3  (a) Temperature dependence of the electrical resistivity of a SIO-214 thin-film on LSGO for current applied along the *b*-axis (red) and *c*-axis (blue).  (b) Arrhenius plot of the resistivity data, where the dashed lines are fits to $\rho\,(T) = \rho_0 \exp\,(\Delta/2k_B T)$ for two distinct temperature regions.

FIG. 4  (a) The optical absorption coefficient for SIO-214 thin films on LSGO where the incident light is polarized such that E⊥*c* (red) and E//*c* (blue).  Schematic band structure and optical transitions (arrows) for (b) the E⊥*c* polarization and (c) the E//*c* polarization.  The dotted lines indicate the Fermi energy.

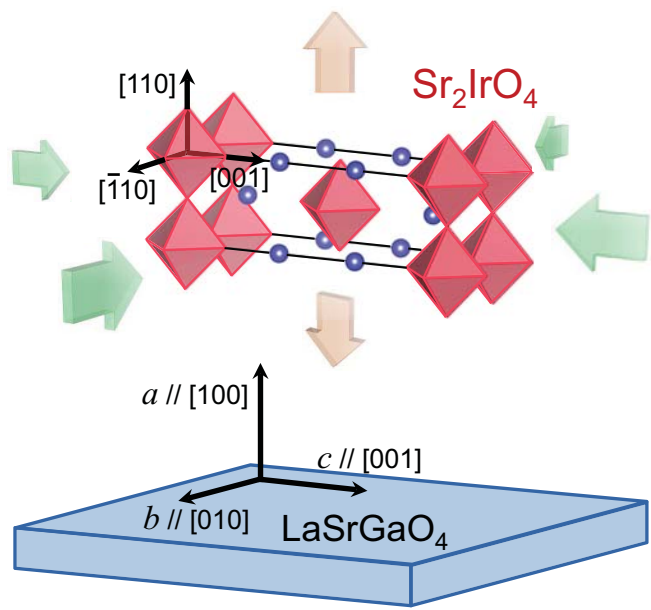

[110]

[$\bar{1}$10]  [001]

Sr$_2$IrO$_4$

$a$ // [100]

$c$ // [001]

$b$ // [010]  LaSrGaO$_4$

Figure 1

Nichols *et al.*

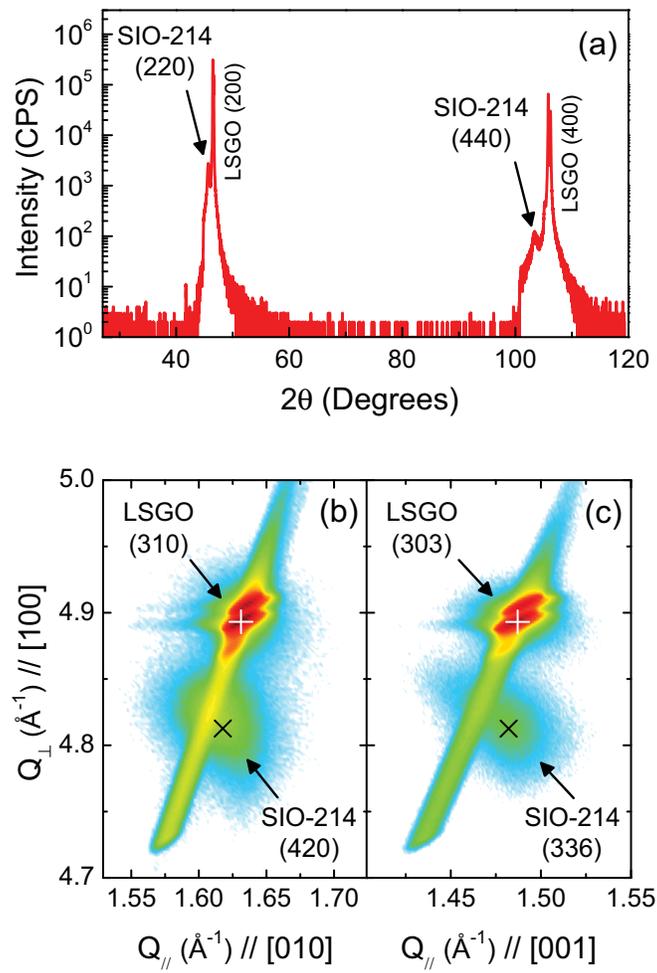

Figure 2

Nichols *et al.*

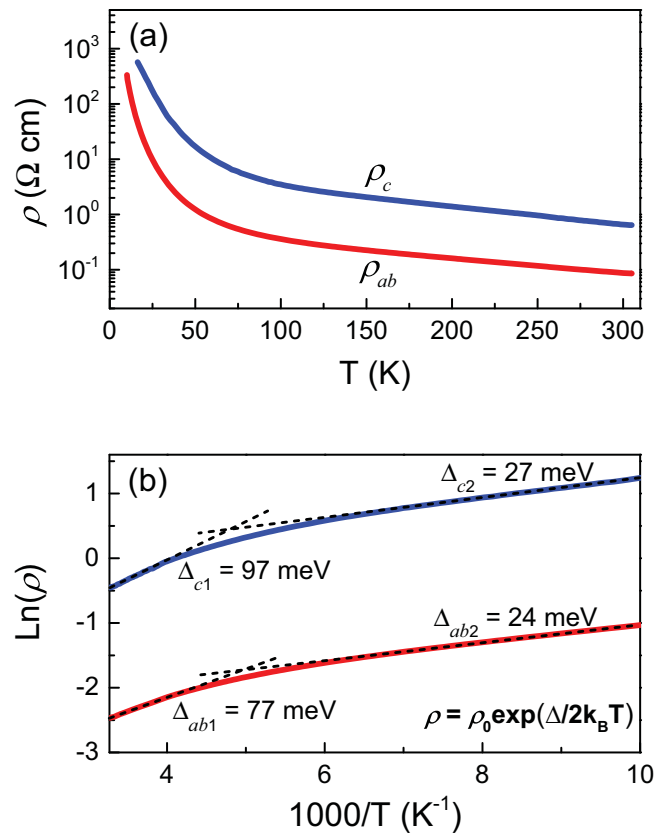

(a)

$\rho_c$

$\rho_{ab}$

(b)

$\Delta_{c2}$ = 27 meV

$\Delta_{c1}$ = 97 meV

$\Delta_{ab2}$ = 24 meV

$\Delta_{ab1}$ = 77 meV

$\rho = \rho_0 \exp(\Delta/2k_BT)$

Figure 3

Nichols *et al.*

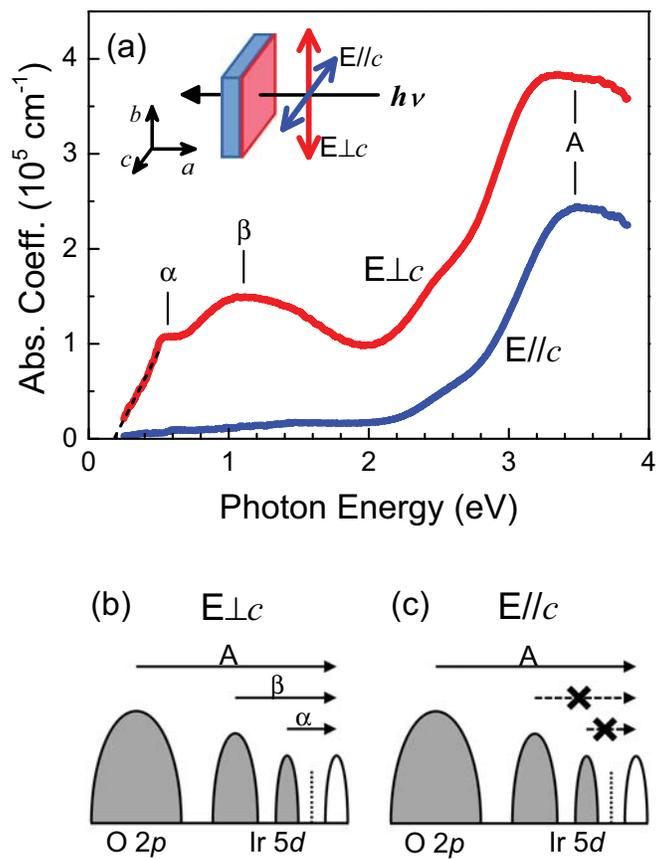

Figure 4

Nichols *et al.*